\documentclass[
 amsmath,amssymb,
 aps,
prl,nofootinbib,twocolumn,superscriptaddress,preprintnumbers,balancelastpage,longbibliography
]{revtex4-2}

\usepackage{amsmath}
\usepackage{graphicx}
\usepackage{dcolumn}
\usepackage{bm}
\usepackage[colorlinks=true,linkcolor=blue,citecolor=blue,urlcolor=blue]{hyperref}

\usepackage[normalem]{ulem}
\usepackage{color}

\def\beq{\begin{equation}}
\def\eeq{\end{equation}}
\newcommand{\anorm}[3]
{\left(\frac{#1}{#2}\right)^{#3}}

\newcommand{\eV}{{\, {\rm eV}}}

\newcommand{\MeV}{{\, {\rm MeV}}}
\newcommand{\GeV}{{\, {\rm GeV}}}
\newcommand{\TeV}{{\, {\rm TeV}}}

\begin{document}

\title{
Pre-inflationary QCD axion stars after moduli domination
}

\author{Edward Hardy}
\email{edward.hardy@physics.ox.ac.uk}
\affiliation{%
 Rudolf Peierls Centre for Theoretical Physics, Department of Physics, University of Oxford, Parks Road, OX1 3PU, UK 
}%
\author{Noelia Sánchez González}
\email{noelia.sanchezgonzalez@physics.ox.ac.uk}
\affiliation{%
 Rudolf Peierls Centre for Theoretical Physics, Department of Physics, University of Oxford, Parks Road, OX1 3PU, UK 
}%
\author{Henry Stubbs}
\email{henry.stubbs@physics.ox.ac.uk}
\affiliation{%
 Rudolf Peierls Centre for Theoretical Physics, Department of Physics, University of Oxford, Parks Road, OX1 3PU, UK 
}
\author{Lorenzo Tranchedone}
\email{lorenzo.tranchedone@physics.ox.ac.uk}
\affiliation{%
 Rudolf Peierls Centre for Theoretical Physics, Department of Physics, University of Oxford, Parks Road, OX1 3PU, UK 
}

\begin{abstract}

The growth of adiabatic density perturbations during an era of early matter domination induces $\mathcal{O}(1)$ fluctuations in pre-inflationary QCD axion dark matter across a broad, string-theory-motivated parameter space. Remarkably, at $\Lambda$CDM matter-radiation equality the scale of these perturbations coincides with the quantum Jeans scale, so they collapse to solitonic ``axion stars''. These axion stars have densities up to $10^4\,\mathrm{eV}^4$, and, including their surrounding halos, they contain as much as $50\%$ of dark matter. Direct searches for a smooth axion background can be suppressed, but transient enhancements or indirect astrophysical signals at axion masses $m_a\lesssim 10^{-5}\,{\rm eV}$ would point to a non-standard cosmological history.

\end{abstract}

\maketitle



{\bf Introduction---}The QCD axion is a well-motivated extension of the Standard Model of particle physics that solves the strong CP problem \cite{Peccei:1977hh,Wilczek:1977pj,Weinberg:1977ma}. Moreover, candidate axions are generic in string theory \cite{Svrcek:2006yi,Conlon:2006tq,Arvanitaki:2009fg,Cicoli:2012sz} and a QCD axion inevitably forms at least a component of dark matter.

There are two classes of axion cosmology, known as the pre- and post-inflationary scenarios \cite{Sikivie:2006ni}. We focus on the pre-inflationary case, where the Peccei-Quinn (PQ) symmetry is broken during inflation and not subsequently restored. This is natural for axions arising from $p$-form fields integrated over compactification cycles, since these constructions have no 4D phase in which the PQ symmetry is unbroken \cite{March-Russell:2021zfq,Benabou:2023npn,Benabou:2025kgx}. 
Such realisations are attractive because the axion shift symmetry can be protected up to exponentially suppressed effects (similar protection can occur in some post-inflationary theories \cite{Petrossian-Byrne:2025mto,Loladze:2025uvf}).

In the pre-inflationary scenario, the axion field is nearly homogeneous over the observable Universe at the end of inflation. Axion dark matter is produced by the misalignment mechanism, with the abundance depending on the initial misalignment angle $\theta_0$ \cite{Preskill:1982cy,Abbott:1982af,Dine:1982ah}. 
We assume that the QCD axion makes up all the dark matter. In a standard cosmological history, this corresponds to an axion decay constant $f_a\simeq 10^{12}\,{\rm GeV}$ 
for $\theta_0\simeq 1$ \cite{GrillidiCortona:2015jxo,Borsanyi:2016ksw}. 

However, observations only require a standard cosmological history below temperatures of about $5\,{\rm MeV}$ \cite{deSalas:2015glj,Hasegawa:2019jsa}, shortly before Big Bang Nucleosynthesis (BBN), and earlier epochs are largely unconstrained. 

String theory compactifications contain scalar fields, known as moduli, that control the size and shape of the extra dimensions.  Also produced by misalignment, these typically come to dominate the energy density of the early Universe, leading to an era of early matter domination (EMD) \cite{Ellis:1986zt,Banks:1993en,Kane:2015jia,Cicoli:2023opf}. 
They decay by operators suppressed by the Planck scale $M_{\rm Pl}\equiv G^{-1/2}$ at a rate $\Gamma\simeq  m_\phi^3/M_{\rm Pl}^2$, where $m_\phi$ is the modulus mass. Moduli are often light, e.g. 
$m_\phi\sim 10^2\,{\rm TeV}$, yielding a low reheating temperature
\begin{equation}
T_{\rm rh} \sim  5\,{\rm MeV}\left(\frac{m_\phi}{10^2\,{\rm TeV}} \right)^{3/2}~,
\end{equation}
defined by 
$\Gamma_\phi= \sqrt{\frac{4\pi^3}{45}g_\star(T_{\rm rh})} T_{\rm rh}^2/M_{\rm Pl}$, after which standard radiation domination begins. 

EMD with $T_{\rm rh}\lesssim {\rm GeV}$ reduces the axion relic abundance, allowing larger $f_a$ for a given $\theta_0$ \cite{Lazarides:1990xp,Kawasaki:1995vt,Giudice:2000ex,Visinelli:2009kt,Arias:2021rer}. This is potentially advantageous for string theory models, which often lead to values of $f_a\gtrsim 10^{14}\,{\rm GeV}$ that would otherwise require $\theta_0\ll 1$ \cite{Banks:1996ea,Acharya:2010zx,Broeckel:2021dpz}. But $\theta_0$ is not directly observable, so a discovery of an axion with large $f_a$ would be, at most, a tentative hint of EMD.

In this paper, we show that EMD has another, more robust, observational signature in QCD axion dark matter, in the form of distinctive substructure. Refs.\,\cite{Nelson:2018via,Visinelli:2018wza,Blinov:2019jqc,WileyDeal:2023trg} previously studied axion substructure from EMD, with Ref.\,\cite{Blinov:2019jqc} including wave effects during EMD. We build on these by demonstrating the central role of wave effects both during EMD and around $\Lambda$CDM matter-radiation equality (MRE), and by performing numerical simulations from early times through MRE.\footnote{Substructure from EMD has been studied for other dark matter candidates, and can lead to new observational signals \cite{Erickcek:2011us,Erickcek:2015jza,Blanco:2019eij,Bramante:2024pyc}. However, in many theories, the strongest effects are erased after the modulus decays \cite{Erickcek:2011us,Fan:2014zua,Barenboim:2013gya,Blinov:2021axd,Barenboim:2021swl,Ganjoo:2023fgg}. An axion retains information because of its non-thermal production and wave nature.}

\vspace{0.2cm}
{\bf Axion dark matter and EMD---}We assume that EMD is driven by a single light modulus $\phi$, as in, e.g., the Large Volume Scenario \cite{Balasubramanian:2005zx,Conlon:2005ki,Apers:2024ffe}. Then, during EMD the temperature $T$ decreases with increasing scale factor $R$ as $T\propto R^{-3/8}$ \cite{Scherrer:1984fd,Giudice:2000ex}. 

The axion begins oscillating at $T\simeq T_\star$, defined by $m_a(T_\star)= 3H(T_\star)$, with $H$ the Hubble parameter and $m_a(T)$ the temperature-dependent axion mass. 
For our purposes it suffices to take
\begin{equation} \label{eq:maT}
m_a(T) = \begin{cases}
 m_a & \text{if } T<\Lambda \\
 m_a \left(\Lambda/T  \right)^{\alpha/2} & \text{otherwise}~,
\end{cases}
\end{equation}
with $\Lambda= 100\,{\rm MeV}$, $\alpha= 8$, and $m_a$ the zero-temperature mass \cite{Gross:1980br}. 
Provided that $T_\star\gtrsim 100\,{\rm MeV}$ and  $T_{\rm rh}\lesssim 100\,{\rm MeV}$, the observed relic abundance requires
\beq
f_a \simeq  10^{14}\,\GeV \anorm{10\,\MeV}{T_{\rm rh}}{4/3} \theta_0^{-4/3}~.
\eeq

For $T_{\rm rh}\sim 10\,\MeV$ values of $f_a$ between $10^{14}\,{\rm GeV}$ and $10^{16}\,{\rm GeV}$ are possible with $\theta_0\gtrsim 0.1$. In this range, 
\beq
T_\star \simeq 140\,\MeV\left(\frac{T_{\rm rh}}{10\,\MeV}\right)^{1/4} \anorm{10^{14}\,\GeV}{f_a}{1/8}~,
\eeq
is self-consistently greater than $100\,{\rm MeV}$.


\vspace{0.2cm}
{\bf Structure formation during EMD---}We consider the minimal scenario in which the initial modulus perturbations are the scale-invariant adiabatic fluctuations from inflation, extrapolated to EMD scales.\footnote{While the axion mass is changing, perturbations in the temperature also lead to some axions being excited to finite momentum modes, prior to the times we consider. In Appendix~\ref{app:res} we show that this effect does not significantly modify our conclusions.} 

During matter domination,  sub-horizon  modulus density perturbations  grow as $\delta_{\phi}\propto R$, while super-horizon modes are frozen. For modes that enter the horizon during EMD, the dimensionless linear power spectrum of the modulus is therefore
\beq \label{eq:power}
\Delta^2_\phi(k) \simeq \Delta^2_0 \left(\frac{k}{RH} \right)^4 ~,
\eeq
where $k$ is the comoving wavenumber and $\Delta^2_0=\left(\frac{2}{5}\right)^2A_s$ with $A_s\simeq 2\cdot 10^{-9}$ 
\cite{Planck:2018vyg,Blinov:2019jqc}.

Axions constitute a tiny fraction of the total energy density throughout EMD. Consequently, perturbations in their energy density do not significantly affect the gravitational potential (and their self-interactions are also negligible). Axions do however fall into the gravitational wells of the modulus perturbations.  
Treating the modulus as an external source, non-relativistic sub-horizon axion energy density perturbations in Newtonian gauge $\delta_{a}(\mathbf{k})\equiv\delta_{a,k}$ evolve as 
\begin{align} \label{eq:deltaa_ev}
    \delta_{a,k}'' +  \mathcal{H}\delta_{a,k}' +  \left(c_{s,a}^2k^2\delta_{a,k}  - 4\pi G R^2 \bar{\rho}_{\rm tot} \delta_{\phi,k}\right) = 0\,,
\end{align}
where $'$ denotes a derivative with respect to conformal time, $\mathcal{H}=R'/R$, $\bar{\rho}_{\rm tot}$ is the average total energy density, and 
$c_{s,a}= k/(2m_a R)$ is the axion sound speed \cite{Arvanitaki:2019rax,Blinov:2019jqc}. Eq.\,\eqref{eq:deltaa_ev} follows from the conservation of the energy-momentum tensor, or equivalently from the non-relativistic limit of the axion field equation of motion after a Madelung transformation (see  Appendix~\ref{app:lin}).

The non-zero $c_{s,a}$ reflects the axion's wave nature (the analogous effects for the modulus are negligible on the scales of interest). 
Balancing wave (quantum) pressure against gravity in Eq.\,\eqref{eq:deltaa_ev} 
identifies an effective axion Jeans scale in the background of the modulus:
\begin{equation}
    k_{J}^{(\rm phys) }=\frac{k_{J}}{R} \equiv \left[16\pi G\;\bar{\rho}_{\rm tot}  m_a^2 \left(\frac{\delta_{\phi,k}}{\delta_{a,k}}\right)\right]^{1/4}~.
\end{equation}
Axion modes at $k<k_{J}$ grow, while those at $k>k_{J}$ oscillate. Unusually, $k_{J}$ depends on the ratio $\delta_{\phi,k}/\delta_{a,k}$. This is because a larger perturbation in the modulus density sources a deeper gravitational well, allowing a greater axion density to build up before its quantum pressure matches the gravitational force.

When $T\gg100\,{\rm MeV}$ the axion mass is suppressed, so the Jeans scale is small and rapidly increasing, and the relevant modes grow negligibly. 
Once the mass saturates, the evolution is governed by a critical scale
\beq \label{eq:kcrit1}
k_{{\rm crit}}\equiv R\left(16\pi G\;\bar{\rho}_{\rm tot}  m_a^2\right)^{1/4}~,
\eeq
which grows as $R^{1/4}$. 
Modes at $k<k_{{\rm crit}}$ reach $\delta_a \simeq\delta_\phi$, so the axion density power spectrum $\Delta^2_a(k)$ tracks $\Delta^2_\phi(k)$. Meanwhile, modes at $k>k_{{\rm crit}}$ 
are limited to $\delta_{a,k}/\delta_{\phi,k} \simeq \left(k_{{\rm crit}}/k\right)^4$ by the Jeans barrier.

\begin{figure}[t!]
    \centering    \includegraphics[width=1.0\columnwidth]{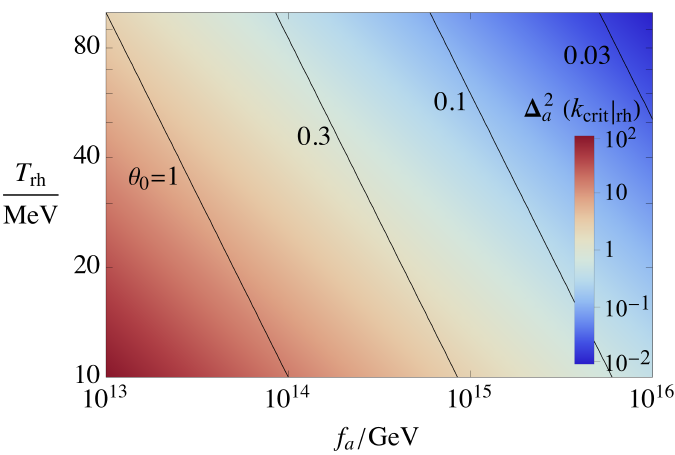}
    \caption{ The peak amplitude of the axion linear power spectrum at the end of EMD  $\Delta_a^2(k_{\rm crit}|_{\rm rh})$. The misalignment angle $\theta_0$ for the axion to comprise the full dark matter abundance is also shown.}
    \label{fig:plot1}
\end{figure}

Given that $\Delta^2_\phi \propto k^4$, we obtain
\begin{equation}
\Delta^2_a(k) \sim \begin{cases}
 \Delta^2_0 \left(k/(RH) \right)^4 & \text{if } k<k_{{\rm crit}}(t) \\
  \Delta^2_0 \left(k_{\rm crit}/(RH) \right)^4 \left( k_{{\rm crit}}/k  \right)^{4} & \text{otherwise}~.
\end{cases}
\end{equation}
The axion power at reheating peaks at $k_{{\rm crit}}|_{\rm rh}$, with 
\beq \label{eq:Dmax}
\begin{aligned}
\Delta^2_a(k_{{\rm crit}}|_{\rm rh}) &\sim \Delta^2_\phi(k_{{\rm crit}}|_{\rm rh}) \\
&\sim 400 \anorm{10\,\MeV}{T_{\rm rh}}{4} \anorm{10^{14}\,\GeV}{f_a}{2}.
\end{aligned}
\eeq
Consequently, as shown in Fig.\,\ref{fig:plot1}, broad classes of EMD scenarios lead to large inhomogeneities in the axion field.

If $\phi$ perturbations on the scale $k_{\rm crit}$ go nonlinear, some axions become gravitationally bound to the resulting collapsed structures (simulations show that structures at smaller spatial scales have little effect on the axion power spectrum, as the axion's de Broglie wavelength would exceed the object's size). These axions 
seed power on smaller comoving length scales. For some $(f_a,T_{\rm rh})$ modes slightly below $k_{\rm crit}$ also reach $\Delta_a^2(k)\sim 1$.


\vspace{0.2cm}
{\bf To MRE---}As the modulus decays at the end of EMD the Universe becomes radiation dominated and the gravitational potential decreases. Axions that were not bound to modulus fluctuations coast with the redshifting peculiar velocities acquired during EMD. Meanwhile, axions that were bound to collapsed modulus structures become unbound and free-stream, while retaining their comoving momentum \cite{Visinelli:2018wza}.

Numerical simulations, which we discuss below, show that if $\Delta^2_a(k_{\rm crit}|_{\rm rh})\lesssim 1$ at reheating, the axion power spectrum remains approximately unchanged and retains a peak near $k_{\rm crit}|_{\rm rh}$. If $\Delta^2_a(k_{\rm crit}|_{\rm rh})\gtrsim 1$ at reheating, the axion power spectrum relaxes to have a peak of order unity. Provided $\Delta^2_a(k_{\rm crit}|_{\rm rh})$ is not many orders of magnitude larger than one, the peak is within a factor of a few of $k_{\rm crit}|_{\rm rh}$. This reflects the characteristic momentum scale of the axions at reheating: since the axion energy density is quadratic in the field, it inherits structure on comparable scales.

The axions then evolve freely until MRE, when their density perturbations begin to dominate. The axion quantum Jeans scale at this time is $k_{J}|_{\rm mre}=R_{\rm mre} \left(16\pi G\bar{\rho}_{\rm mre}m_a^2 \right)^{1/4}$. Crucially, this parametrically coincides with the peak of the axion power spectrum:
\beq
\begin{aligned} \label{eq:key}
\frac{k_{\rm crit}|_{\rm rh}}{k_{J}|_{\rm mre}} &= \frac{R_{\rm rh} \left(16\pi G\bar{\rho}_{\rm tot, rh}m_a^2 \right)^{1/4}}{R_{\rm mre} \left(16\pi G\bar{\rho}_{\rm mre}m_a^2 \right)^{1/4}} \\
&\simeq \frac{R_{\rm rh} T_{\rm rh}}{R_{\rm mre} T_{\rm mre}} \simeq 1~,
\end{aligned}
\eeq
where we used $RT=\mathrm{const.}$ during radiation domination, and neglected corrections from changes in the number of relativistic degrees of freedom and transitions between epochs 
(these modify Eq.\,\eqref{eq:key} by a factor less than $1.5$).


\vspace{0.2cm}
{\bf Axion stars---}The largest axion density contrast perturbations, which are typically $\mathcal{O}(1)$, therefore lie at the boundary between gravitational collapse and quantum pressure support at MRE. These form axion stars: self-gravitating solitonic configurations in which gravity is balanced by quantum pressure rather than virial motion.

Ground-state axion stars are spherically symmetric with density profile
\beq
\rho(r)=\rho_c \,\chi\left(\frac{r}{\lambda_J(\rho_c)} \right)~,
\eeq
where $\rho_c$ is the central density, $\lambda_J(\rho_c)=2\pi/k_J(\rho_c)$, and $\chi$ is a universal dimensionless profile with a characteristic core \cite{Ruffini:1969qy,Membrado:1989bqo,Chavanis:2011zm,Visinelli:2017ooc}. An axion star's mass $M$ and half-mass radius $R_{1/2}$ satisfy $MR_{1/2}\simeq 3.9/(Gm_a^2)$ and $\rho_c\simeq G^3m_a^6M^4/(64\pi)$, with the axion's de Broglie wavelength of order $R_{1/2}$. 

The stars are produced with excited quasinormal modes. Also, they are surrounded by halos, where both quantum pressure and velocity dispersion are important and the profile deviates from a pure soliton \cite{Gorghetto:2022sue,Gorghetto:2024vnp}.

Overdensities with $\delta_{a}>1$ collapse at $R/R_{\rm mre}\sim 1/\delta_{a}$, producing axion stars with $\rho_c\simeq \bar{\rho}_{\rm mre} (1+\delta_{a})\delta_{a}^3 \sim (1\textup{--} 10^4)\,{\rm eV}^4$ \cite{Kolb:1993zz}. The masses of stars from perturbations on a scale $k_\delta$ are
\begin{align} 
M &\simeq (1+\delta_{a}) \left(\frac{k_{J}|_{\rm mre}}{k_\delta}\right)^3 M_{J,{\rm mre}}~, \label{eq:masses} \\
\llap{\text{with}\quad}
M_{J,{\rm mre}} &=\lambda_J(\bar{\rho}_{\rm mre})^3 \bar{\rho}_{\rm mre} \notag\\
&\simeq 2\cdot 10^{-13}M_\odot \anorm{f_a}{10^{14}\,\GeV}{3/2}~. \label{eq:masses2}
\end{align}
Since $k_\delta\simeq k_{J}|_{\rm mre}$, these values lie well below the maximum stable axion star mass from the quartic interaction, $10^{-11}M_\odot f_a/(10^{14}\,\GeV)$  \cite{Chavanis:2022fvh}.

After MRE, $k_{J}$ grows allowing smaller-scale overdensities to collapse into lower-mass, lower-density stars. 
Initially perturbative modes with $k<k_{J}|_{\rm mre}$ also grow and collapse. 
The resulting structures increasingly deviate from the solitonic solution. Some early-formed stars are bound inside larger structures, in a process analogous to hierarchical structure formation, and later merge into halos sourced by modes that were super-horizon during EMD, including, eventually, galaxies.


\vspace{0.2cm}
{\bf Simulations---}To test the analysis above and determine numerical coefficients, we perform simulations from EMD through MRE. The key results are presented here, with further details in Appendix~\ref{app:sims}.

We solve the evolution of the axion and modulus fields from $T=100\,\MeV$ on a periodic lattice in a cosmological background. The box size and resolution are such that all modes are sub-horizon and non-relativistic, reducing the dynamics to the Schr\"odinger–Poisson system. The modulus is initialised as a Gaussian random field with  power spectrum of Eq.\,\eqref{eq:power}, while the axion is initially homogeneous. 
Modulus decay is modelled by uniformly reducing its energy density and transferring this to a homogeneous radiation component. 
The lattice UV scale is taken well above $k_{\rm crit}|_{\rm rh}$. To control UV artefacts, we use a reduced modulus mass, such that its Jeans scale lies above $k_{\rm crit}|_{\rm rh}$ throughout. We have verified that systematic uncertainties from this modification, as well as the finite box size, timestep, and lattice spacing, are negligible.

During EMD,  $\Delta^2_a(k)\simeq \Delta^2_\phi(k)$ for $k\lesssim k_{\rm crit}$, while power at $k\gtrsim k_{\rm crit}$ is initially suppressed but grows if modes around $k_{\rm crit}$ become nonlinear. After reheating, $\Delta^2_a$ develops the expected peak near $k_{\rm crit}|_{\rm rh}\simeq k_{J}|_{\rm mre}$, both in the perturbative and nonlinear cases. 
For $T_{\rm rh}=20\,\MeV$, the peak is at $k\simeq 1.5\,k_{J}|_{\rm mre}$ for  $f_a=10^{15}\,\GeV$. It shifts to smaller $k/k_{J}|_{\rm mre}$ for larger $f_a$ ($0.5\,k_{J}|_{\rm mre}$ for $f_a=10^{16}\,\GeV$) and to larger $k$ for smaller $f_a$ ($3\,k_{J}|_{\rm mre}$ for $f_a=3\cdot10^{14}\,\GeV$), corresponding to more linear and more nonlinear evolution respectively.

\begin{figure}[t!]
    \centering    \includegraphics[width=1.0\columnwidth]{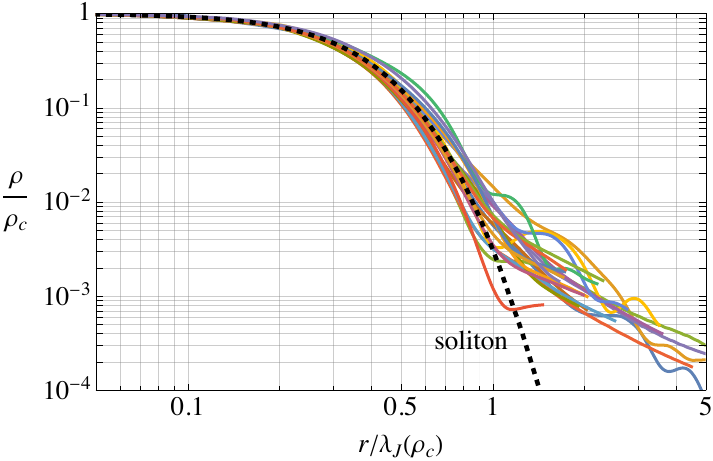}
    \caption{Spherically averaged density profiles of the densest objects at MRE in an example simulation (coloured), compared to the solitonic axion star prediction (dashed black).}
    \label{fig:profs}
\end{figure}

During radiation domination the axion power spectrum is nearly constant. Shortly before MRE, bound structures form, indicated by the decoupling of the densest regions from Hubble expansion and the growth of small-scale power.

In Fig.\,\ref{fig:profs} we show the spherically averaged density profiles of the  $20$ densest objects at $R=R_{\rm mre}$ for $f_a=10^{15}\,\GeV$ with $T_{\rm rh}=20\,\MeV$ in a typical simulation. 
Each profile is rescaled by its central density $\rho_c$ and $\lambda_J(\rho_c)$, so that solitonic configurations take a universal form. For $r\lesssim \lambda_J(\rho_c)$, the profiles closely follow this, indicating substantial axion star cores. At larger radii they deviate, resembling halos but with the axion de Broglie wavelength comparable to their size. In the central regions, the radial component of the quantum pressure balances the gravitational potential gradient.

\begin{figure}[t!]
    \centering    \includegraphics[width=1.0\columnwidth]{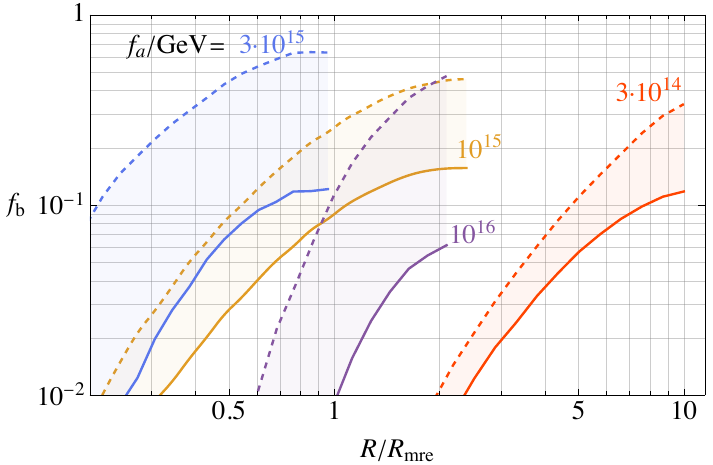}
    \caption{
    The fraction $f_b$ of axions bound in axion stars (solid) and in stars and their halos (dashed) as a function of the scale factor for different $f_a$, with $T_{\rm rh}=20\,\MeV$.}
    \label{fig:frac}
\end{figure}

Fig.\,\ref{fig:frac} shows the fraction of axions in stars, and in stars plus surrounding halos, for different $f_a$, with $T_{\rm rh}=20\,\MeV$. Stars are defined as regions whose profiles match the soliton form within a factor of $1.25$ 
(relaxing the agreement criterion to $1.5$ increases the fraction of axions in stars by roughly a factor of $1.3$). 
Halos are taken to extend to $5\bar{\rho}_a$, which is approximate since their edges are not sharply defined. Simulations end once the densest solitons are no longer resolved.

The fraction of axions in stars exceeds $15\%$ for $f_a=10^{15}\,\GeV$, with most forming around MRE. 
The densest stars appear first, followed by less-dense, lower-mass ones. The early stars have masses a few times $M_{J,{\rm mre}}$ and densities up to $10^5\bar{\rho}_{\rm mre}$, while later stars are around $M_{J,{\rm mre}}$. The total bound fraction reaches $50\%$.

The qualitative behaviour is similar for $3\cdot10^{14}\,\GeV\lesssim f_a\lesssim10^{16}\,\GeV$ with $T_{\rm rh}=20\,\MeV$, corresponding to $\Delta^2_a(k_{\rm crit}|_{\rm rh})$ from less than one to nonlinear at reheating. For $f_a$ near the lower end of this range, the peak is shifted to slightly larger $k/k_{J}|_{\rm mre}$, delaying formation of stars until a little after MRE. For $f_a$ towards the upper end, the spectrum is slightly more infrared so the bound objects are further from isolated solitons (with soliton fraction of about $8\%$ for $f_a=10^{16}\,\GeV$). In all cases, the total bound fraction exceeds $30\%$. Smaller $f_a$ cannot be probed due to too much power reaching the lattice scale. More generally, the properties of the bound structures are controlled by $\Delta^2_a(k_{\rm crit}|_{\rm rh})$, with different $(f_a,T_{\rm rh})$ giving similar results when this is fixed.

\begin{figure}[t!]
    \centering    \includegraphics[width=1.0\columnwidth]{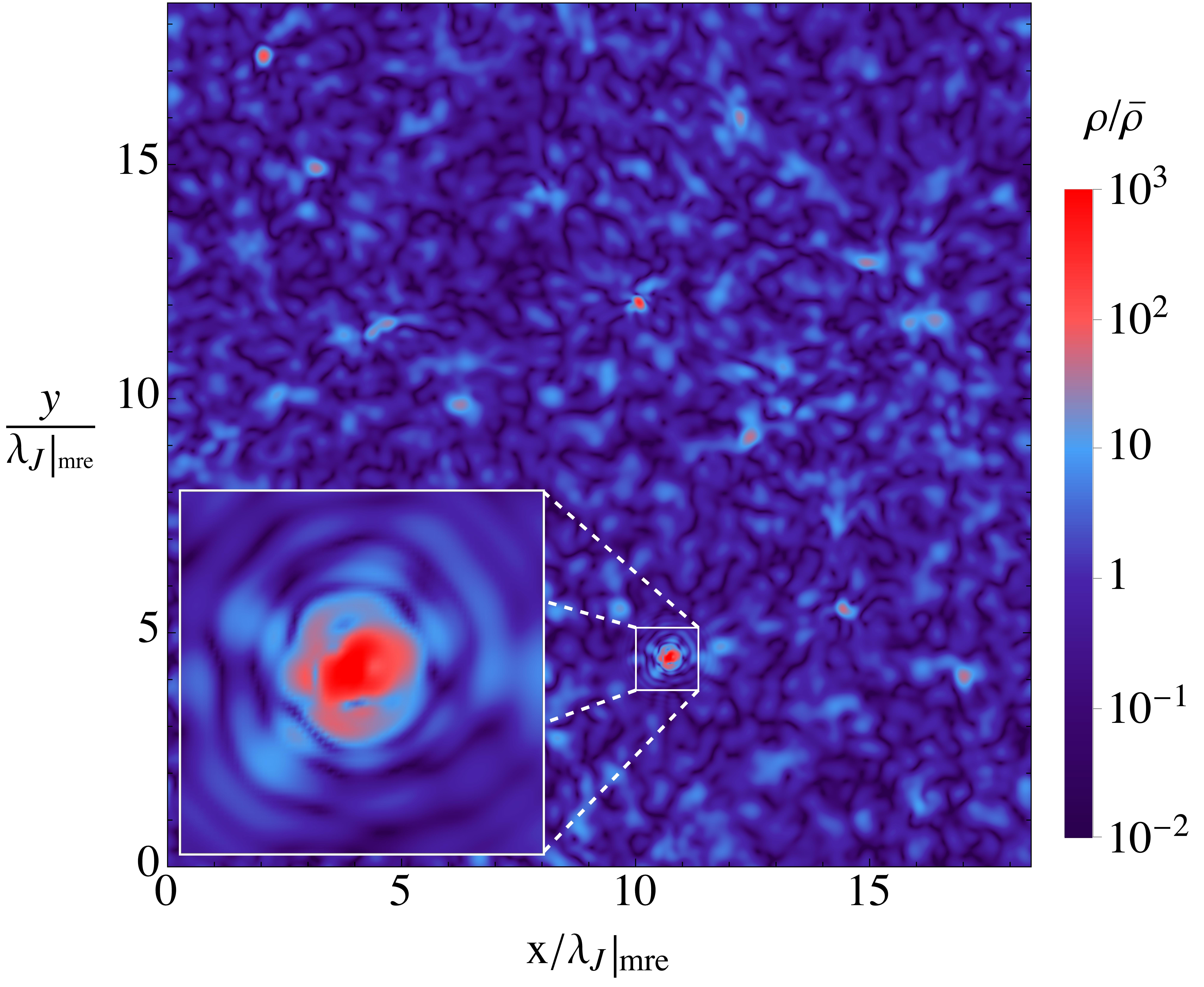}
    \caption{The axion energy density through a slice of a simulation, relative to the mean density $\bar{\rho}$, at $R/R_{\rm mre}=2$ for $f_a=10^{15}\,\GeV$ and $T_{\rm rh}=20\,\MeV$.}
    \label{fig:slice}
\end{figure}

A slice of the axion density soon after MRE is plotted in Fig.\,\ref{fig:slice}, showing a typical star and halo.


\vspace{0.2cm}
{\bf Discussion---}A key open question is the subsequent evolution of axion stars and their host halos. These may grow through accretion and mergers, potentially approaching the stability limit, or be disrupted as they are incorporated into larger structures \cite{Green:2006hh,vandenBosch:2017ynq,Delos:2019tsl,Shen:2022ltx}. Analytic estimates for post-inflationary axions suggest the solitons that form around MRE are relatively robust \cite{Dokuchaev:2017psd}, but dedicated simulations are required. Improved numerical methods, such as adaptive meshes, will be important for resolving these dynamics, as well as probing regimes with $\Delta^2_a(k_{\rm crit}|_{\rm rh})\gg 1$.

Our results have implications for direct detection searches for axions with $m_a\lesssim 10^{-5}\,\eV$, where EMD is best motivated, e.g., \cite{Kahn:2016aff,Ankel:2026zrv}. 
If an $\mathcal{O}(1)$ fraction of axions are bound in substructure, and this survives disruption in the Solar neighbourhood, the smooth background is reduced, weakening experimental signals. 
The densest axion stars, which have masses around $M_{J,{\rm mre}}$ defined in Eq.\,\eqref{eq:masses2}, have radii
\beq
R_{1/2} \simeq 1.6\cdot 10^{11}\,{\rm m}\left(\frac{f_a}{10^{14}\,\GeV} \right)^{1/2} \frac{M_{J,{\rm mre}}}{M}\,.
\eeq
Collisions between the Earth and these stars 
are extremely rare, with the typical interval between events
\beq
t_{\rm coll} \simeq 10^4\,{\rm yr}\left(\frac{f_a}{10^{14}\,\GeV}\right)^{1/2}\,,
\eeq
assuming $20\%$ of dark matter is in objects of mass $0.1M_{J,{\rm mre}}$. More frequent transient enhancements may arise from larger-scale modes ($k<k_{\rm crit}|_{\rm rh}$) or streams from disrupted objects  \cite{Tinyakov:2015cgg}, motivating broadband searches \cite{Kahn:2016aff}. 
New indirect signals may also arise \cite{Kolb:1995bu,Tkachev:2014dpa,Braaten:2016dlp,Fairbairn:2017sil,Fairbairn:2017dmf,Hertzberg:2018zte,Du:2023jxh,Escudero:2023vgv,Fox:2023xgx,Xiao:2024qay,Chang:2024fol,Fox:2025tqa}, e.g., photon emission during axion star–neutron star encounters \cite{Eby:2017xaw,Bai:2017feq} or from caustic microlensing \cite{Dai:2019lud}. 
Although we focused on the QCD axion, similar conclusions apply to axion-like particles with EMD. It would also be interesting to explore the impact of non-gravitational couplings between the axion and modulus, as may arise in specific string compactifications.

Axion substructure also arises in the post-inflationary scenario \cite{Hogan:1988mp,Kolb:1993zz,Vaquero:2018tib,Eggemeier:2019khm,Xiao:2021nkb}. In this case, the axion power spectrum, relative to $k_{J}|_{\rm mre}$, is similar to the case we have considered, leading to a comparable population of axion stars \cite{Gorghetto:2024vnp}. However, the axion masses in our scenario are much smaller than in the post-inflationary case in the absence of EMD  ($m_a\gtrsim 10^{-4}\text{--}10^{-3}\,\eV$) \cite{Gorghetto:2018myk,Gorghetto:2020qws,Buschmann:2021sdq,Saikawa:2024bta}. In a standard cosmological history pre-inflationary axions produce substructure only for $\theta_0\simeq \pi$, corresponding to large $m_a$ \cite{Arvanitaki:2019rax}. 
A discovery of QCD axion stars at $m_a\lesssim 10^{-5}\,\eV$ would therefore provide strong evidence for EMD.  Conversely, a discovery of a QCD axion without substructure would constrain the early evolution of the Universe beyond current bounds from BBN.

\vspace{0.25cm}
\begin{acknowledgments}
{\bf Acknowledgements}

We thank Joseph Conlon for useful discussions, and EH thanks Marco Gorghetto and Giovanni Villadoro for collaboration on related work. 
EH acknowledges the UK Research and Innovation Future Leader Fellowship MR/V024566/1. 
NSG acknowledges support from the Oxford-Berman Graduate Scholarship jointly funded by the Clarendon Fund and the Rudolf Peierls Centre for Theoretical
Physics Studentship. 
HS is supported by The Science and Technology Facilities Council under the grant ST/X508664/1. 
LT is supported by the Science and Technology Facilities Council Doctoral Training Partnership under the grant ST/Y509474/1. 
For the purpose of
open access, the author has applied a CC BY public copyright license to any Author Accepted
Manuscript (AAM) version arising from this submission.

\end{acknowledgments}

\appendix

\section{Resonances during axion mass turn-on}\label{app:res}

Spatial inhomogeneities in the plasma temperature arising from adiabatic perturbations  
can resonantly excite high-momentum axion modes 
during the period when the axion mass is changing, $T\gtrsim 100\,\MeV$ \cite{Sikivie:2021trt,Kitajima:2021inh,Ayad:2025awu}. 
In a standard cosmological history this effect is too weak to generate significant axion substructure  unless the axion is initially very close to the top of its potential with misalignment angle $\theta_0 \gtrsim \pi- 10^{-3}$. Here we extend the analysis of Ref.\,\cite{Kitajima:2021inh} to EMD and show that the resonant effect remains negligible.

The homogeneous mode of the rescaled axion field $\theta \equiv a/f_a$ satisfies the equation of motion
\beq\label{eq:appzeromode}
\theta'' + 2\mathcal{H} \theta' +R^2 m_a^2(T) \sin{\theta} = 0 \, .
\eeq
Unlike the main text, we parametrise the temperature dependence of the axion mass with the smooth function
\beq
m_a^2(T) = \frac{m_a^2}{1 + (T/\Lambda)^\alpha}\, .
\eeq
where $\Lambda$ and $\alpha$ are defined as in Eq.\,\eqref{eq:maT}.  
We work in conformal time $\eta$, with units such that $\eta = 1$, $R = R_c$ and $H = H_c$ when $T = \Lambda$. We also assume pure modulus domination for $T>T_{\rm rh}$, such that $R\sim\eta^2$ and $T \sim R^{-3/8}$ throughout this time. The linear Fourier space perturbation mode of the axion field $\delta \theta_k$ evolves according to
\beq
\begin{aligned}\label{eq:appdeltafull}
    \delta\theta_k'' +\frac{4}{\eta} \delta\theta_k' +  \left[k^2 + R^2 m_a^2(T) \cos{\theta} \right] \delta\theta_k =  j_k \, ,
\end{aligned}
\eeq
where the source term $j_k$ has the form
\beq\label{eq:appsource}
 j_k = R^2 m_a^2(T) \sin{\theta}\left(2 \Phi_k +  \frac{\alpha}{1+\eta^{3\alpha/4}}\frac{\delta T_k}{T} \right)\, .
\eeq
The first term corresponds to the growth due to axions falling onto the perturbations of the gravitational potential $\Phi_k$, while the second term is associated with the effect of temperature inhomogeneities.

Temperature inhomogeneities are related to density perturbations in the radiation bath by 
\beq
    \frac{\delta T_k}{T} = \frac{1}{4}\frac{\delta \rho_{r,k}}{\bar{\rho}_r} \equiv \frac{1}{4} \delta_{r,k}\,.
\eeq
The evolution of $\delta_{r,k}\equiv\delta_r(\mathbf{k})$ sourced by a decaying modulus has been studied in \cite{Erickcek:2011us,Fan:2014zua}. Sub-horizon density perturbations in radiation initially grow together with the modulus; however, unlike the modulus, radiation inhomogeneities quickly saturate and oscillate with constant amplitude due to pressure. Assuming adiabatic initial conditions $\delta_{r,k}^{(0)} = -\Phi_k$, the maximum value of $\delta_{r,k}$ can be obtained numerically (see e.g. Ref.\,\cite{Fan:2014zua}):
\begin{equation}\label{eq:temperaturespec}
 \frac{\delta T_k}{T}\simeq \begin{cases}
\Phi_k/4 & \text{if } k/aH< 1 \\
 9\Phi_k/4 & \text{otherwise}~.
\end{cases}
\end{equation}
$\Phi_k$ is constant in time on all scales during matter domination\footnote{More precisely, the saturation value can differ from Eq.~\eqref{eq:temperaturespec} by some $\mathcal{O}(1)$ factor, set by the exact value of the ratio $\Gamma_\phi/H_{\rm rh}$. Including this factor does not affect our conclusions.}. As in the main text, we take the typical amplitude of primordial gravitational potential fluctuations to be  $\Phi = 3\cdot 10^{-5}$, as set by quantum fluctuations during inflation, so the linear analysis remains valid until near the end of EMD.

We solve Eqs.~\eqref{eq:appzeromode} and \eqref{eq:appdeltafull} numerically and study the evolution of the axion energy density contrast $\delta_{a,k}$ with and without temperature inhomogeneities. As expected from Eq.\,\eqref{eq:appsource}, at early times the effect of temperature perturbations on the growth of sub-horizon modes dominates over direct gravitational interactions, but it quickly becomes negligible as the axion mass saturates. 

\begin{figure}[t!]
    \centering \includegraphics[width=1.0\columnwidth]{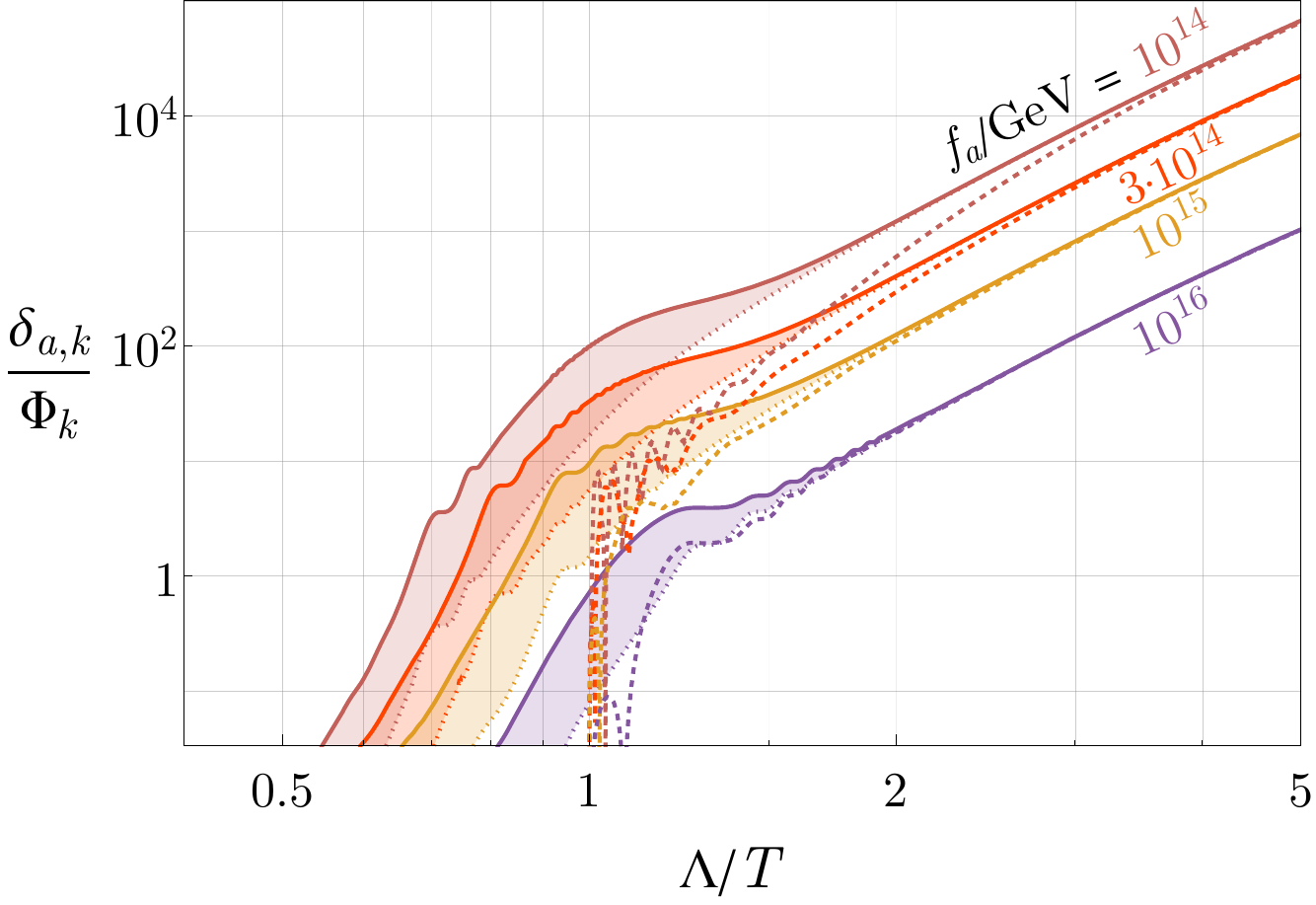}
    \caption{
    Growth of axion density perturbations during axion mass turn-on and the subsequent evolution until reheating, computed in linear perturbation theory and normalised to the gravitational potential perturbation $\Phi_k$ sourcing them. Results are shown for the mode with wavenumber $k = k_{\rm crit}|_{\rm rh}$ for  different $f_a$, with $T_{\rm rh} = 20\,\mathrm{MeV}$ in all cases. Solid curves include the effects of temperature inhomogeneities, while dotted curves are obtained by setting $\delta T_k = 0$. Dashed curves show the evolution with the axion field taken to be homogeneous at $T = \Lambda$ and neglecting temperature inhomogeneities.
    }
    \label{fig:da_resonance}
\end{figure}

In Fig.\,\ref{fig:da_resonance}, we plot the axion energy density perturbations $\delta_{a,k}$ through the axion mass turn-on and up to reheating ($3\Lambda>T>T_{\rm rh}$, where we assume $T_{\rm rh} = 20\,\MeV$), normalised by $\Phi_k$, for modes with $k = k_{\rm crit}|_{{\rm rh}}$ and a range of $f_a$. 
We set the initial misalignment angle $\theta_0=0.2$ in all cases; however, $\delta_{a,k}$ is approximately independent of this choice provided $\theta_0\lesssim 1$, so that $\cos\theta\simeq1$ and $\sin\theta\simeq \theta$ in Eq.\,\eqref{eq:appdeltafull}. 
We show the results obtained including both terms in Eq.\,\eqref{eq:appsource} and only the direct effect of $\Phi_k$. The temperature effects enhance $\delta_{a,k}$ by roughly an order of magnitude when $T=\Lambda$, while at later times their effect becomes negligible as the gravitational collapse catches up. By reheating, the effects from the $\delta T_k$ contribution are completely washed out. 

Figure \ref{fig:da_resonance} also shows the growth of $\delta_{a,k}$ assuming the axion field is homogeneous at $\eta = 1$ and without temperature fluctuations, which is analogous to the evolution implemented in our simulations (see Appendix~\ref{app:sims}). For the range of decay constants probed in the simulations ($f_a \gtrsim 3 \cdot 10^{14} \,\GeV$), the density contrast quickly grows to match the values obtained by the full axion evolution across mass turn-on.

We also verified numerically that temperature inhomogeneities do not have a significant effect at $k<k_{\rm crit}|_{\rm rh}$ or $k>k_{\rm crit}|_{\rm rh}$ and hence do not affect the form of the power spectrum derived in the main text, which remains peaked around $k_{\rm crit}$. In particular, we find that $\delta_{a,k}\sim k^{-2}$ for modes $k \gg k_{\rm crit}|_{\rm rh}$, consistent with $\Delta^2_a(k) \sim k^{-4}$.

We therefore conclude that the resonant excitation can be neglected within the parameter space of interest and does not affect the conclusions of the main text.

\section{Fluid description}\label{app:lin}

To study the effect of quantum pressure on the evolution of axion perturbations, we treat the axion field as a perfect fluid with equation of state $w_a = \bar{P_a}/\bar{\rho_a} \approx0$ and sound speed $c_{s,a}^2 \equiv \delta P_a/\delta \rho_a = k^2/(4m_a^2 R^2)$. This is valid in the non-relativistic limit $k/R\ll m_a$~\cite{Cembranos:2015oya}, assuming that the axion has started to oscillate and that its mass has saturated ($T\lesssim100\MeV$). Conservation of the stress-energy tensor then gives the continuity and Euler equations in Newtonian gauge~\cite{Baumann:2022mni}
\begin{align}
    \delta' &= -(1+w)(\mathbf{\nabla\cdot v}-3\Phi')-3\mathcal{H}(c_s^2-w)\delta,\\
    \mathbf{v}' &= -\mathcal{H}(1-3w)\mathbf{v}-\frac{c_s^2}{1+w}\mathbf{\nabla}\delta-\mathbf{\nabla}\Phi,
\end{align}
where primes denote derivatives with respect to conformal time, $\eta$. Using that $\Phi'\ll\mathbf{\nabla\cdot v}$ in the sub-horizon limit this gives the axion fluid equation of motion
\begin{align} \label{eq: axion fluid eom}
    \delta_{a,k}'' +  \mathcal{H}\delta_{a,k}' + c_{s,a}^2 k^2 \delta_{a,k} = -k^2 \Phi_k 
\end{align}
for the axion fluid. Eq.\,\eqref{eq: axion fluid eom} can equivalently be derived from the axion field equation~\eqref{eq:appdeltafull} (in the case where the mass is temperature independent) by first taking the non-relativistic limit, as described in Appendix~\ref{app:sims}, to give the Schr\"odinger equation~\eqref{eq:Schrodinger Poisson}. Then, carrying out a Madelung transformation by writing the non-relativistic axion field $\psi_{a}=\sqrt{\rho_{a} R^3}e^{i\theta_{a}}$, one obtains a perfect fluid system with density $\rho_{a}$ and velocity $v_{a}=(Rm_{a})^{-1}\nabla\theta_{a}$ (see e.g. \cite{Gorghetto:2022sue}). Working at linear order in perturbations from homogeneity and eliminating $v_a$ finally gives Eq.\,\eqref{eq: axion fluid eom}.

During EMD, the gravitational perturbations are dominantly sourced by the modulus and are determined by Poisson's equation
\begin{align}
    \nabla^2 \Phi = 4\pi G R^2 \bar{\rho}_{\rm tot} \delta_\phi,
\end{align}
so that the equation for the axion perturbations becomes
\begin{align} \label{eq:B6}
    \delta_{a,k}'' +  \mathcal{H}\delta_{a,k}' + c_{s,a}^2 \delta_{a,k}\left(k^2 - \frac{4\pi G R^2 \bar{\rho}_{\rm tot}}{c_{s,a}^2} \frac{\delta_{\phi,k}}{\delta_{a,k}}\right) = 0.
\end{align}
The sign of the final term determines whether the perturbations grow or oscillate, making it useful to introduce the Jeans scale
\begin{equation}
    k_{J}^{(\rm phys) }=\frac{k_{J}}{R} \equiv \left[16\pi G\;\bar{\rho}_{\rm tot}  m_a^2 \left(\frac{\delta_{\phi,k}}{\delta_{a,k}}\right)\right]^{1/4},
\end{equation}
which corresponds to the length scale below which quantum pressure prevents perturbations from growing. As discussed in the main text, $k_J$ differs from the usual expression by a factor of $(\delta_{\phi,k}/\delta_{a,k})^4$, which does not appear in the Jeans scale of the species that dominates the energy density of the Universe, since in that case the gravitational perturbations and the quantum pressure are both sourced by the same overdensity.

\begin{figure}
    \centering
    \includegraphics[width=\columnwidth]{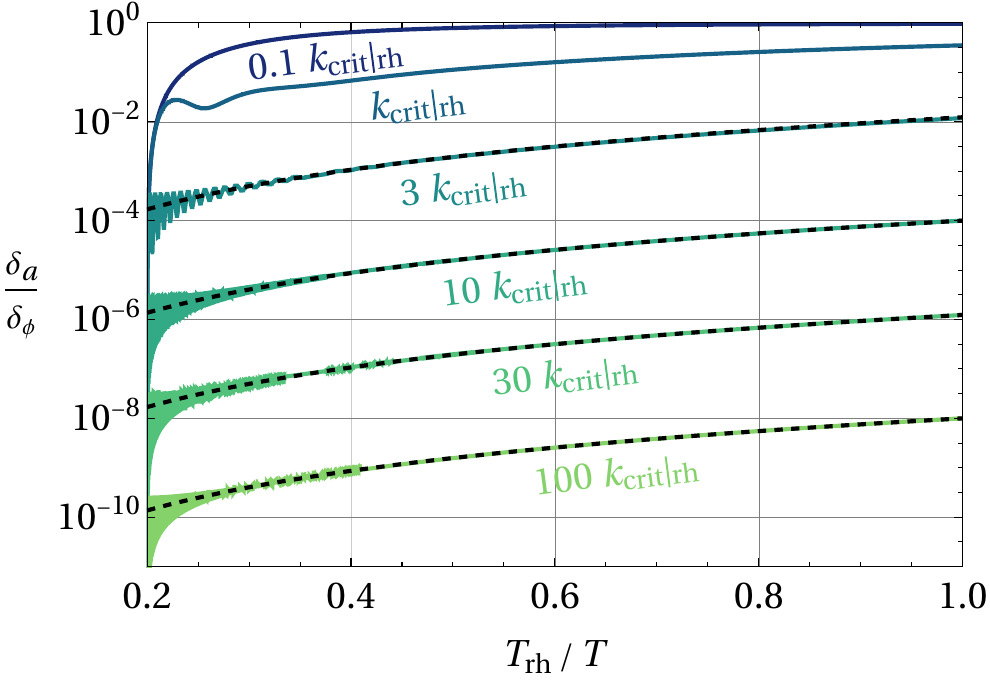}
    \caption{The ratio between the axion and modulus density contrast during EMD, found by numerically solving Eq.\,\eqref{eq: axion fluid eom} for a range of wavenumbers relative to the critical wavenumber at reheating. The black dashed lines show the Jeans barrier prediction that $\delta_a/\delta_\phi = (k_{\rm crit}(T)/k)^4$. In all cases $T_{\rm rh} = 20\,\MeV$ and $f_a=10^{14}\,\GeV$.}
    \label{fig:axion perturbations relative to modulus}
\end{figure}

The ratio between the axion and modulus density contrast, obtained by solving Eq.\,\eqref{eq: axion fluid eom} numerically with an initially homogeneous axion field at $T_{\rm rh}/T=1/5$, is shown in Fig.\,\ref{fig:axion perturbations relative to modulus} for a range of wavenumbers. 
For $k<k_{\rm crit}$, $\delta_{a,k}$ catches up with $\delta_{\phi,k}$ (for super-horizon modes the growth of these adiabatic perturbations is guaranteed by Weinberg's theorem~\cite{Allali:2025pja}). However, for $k>k_{\rm crit}$ the growth is suppressed by the Jeans barrier to 
\begin{equation}
    \frac{\delta_{a,k}}{\delta_{\phi,k}} = \left(\frac{k_{\rm crit}(T)}{k}\right)^4=\left(\frac{k_{\rm crit}|_{\rm rh}}{k}\right)^4\left(\frac{T_{\rm rh}}{T}\right)^{8/3},
\end{equation}
the predictions of which are displayed in dashed black. The initial decaying oscillations are due to a mode which can be found by solving the homogeneous part of Eq.\,\eqref{eq: axion fluid eom}.

\section{Further simulation details}\label{app:sims}

In numerical simulations, we evolve the non-relativistic axion field $\psi_a$, related to the relativistic axion field $a$ by 
\beq \label{eq:NR}
a = \frac{1}{\sqrt{2m_a^2R^3}}\left( \psi_a e^{-i m_a t} +{\rm c.c.}\right)~,
\eeq
and analogously for the non-relativistic modulus field $\psi_\phi$, with $m_\phi$ in place of $m_a$. In the limit $\partial_t \psi_{\phi,a}\ll m_{\phi,a}\psi_{\phi,a}$ and $\partial_t^2 \psi_{\phi,a}\ll m_{\phi,a}^2\psi_{\phi,a}$,
the equations of motion for sub-horizon modes reduce to
\beq \label{eq:Schrodinger Poisson}
\begin{aligned}
&\left( i\partial_t + \frac{\nabla^2}{2m_a R^2} - m_a \Phi\right)\psi_a =0 \\
&\left( i\partial_t + \frac{\nabla^2}{2m_\phi R^2} - m_\phi \Phi\right)\psi_\phi =0 \\
&\nabla^2\Phi = \frac{4\pi G}{R}\left( |\psi_a|^2-\left<|\psi_a|^2\right> + |\psi_\phi|^2-\left<|\psi_\phi|^2\right> \right)~,
\end{aligned}
\eeq
where $t$ is the cosmological time and spatial derivatives are with respect to comoving distances. From Eq.\,\eqref{eq:NR}, the energy densities of the axion and modulus fields are given by $\rho_{a,\phi} =\left| \psi_{a,\phi}\right|^2/R^3$.

The background cosmological expansion is obtained by solving the Friedmann equations
\beq
\frac{\dot{R}}{R}=H = \sqrt{\frac{8\pi G}{3}\rho_{\rm tot}} ~,
\eeq
with total energy density $\rho_{\rm tot}=\rho_{\rm rad}+\rho_a+\rho_\phi$, where $\rho_{\rm rad}$ is a smooth radiation component (we neglect Standard Model matter). The energy densities evolve according to 
\beq \label{eq:energies}
\begin{aligned}
\frac{d\left<\rho_\phi\right>}{dt} &= -\Gamma_\phi \left<\rho_\phi\right> - 3H\left<\rho_\phi\right> \\
\frac{d\left<\rho_r\right>}{dt} &= +\Gamma_\phi \left<\rho_\phi\right> - 4H\left<\rho_r\right> \\
\frac{d\left<\rho_a\right>}{dt} &= - 3H\left<\rho_a\right> ~,
\end{aligned}
\eeq
where $\Gamma_\phi$ is the modulus decay rate to radiation. Given Eq.\,\eqref{eq:NR},  $\left<|\psi_a|^2\right>$ remains constant in time. 

To account for modulus decay, which is not included in Eq.\,\eqref{eq:Schrodinger Poisson}, we uniformly rescale $\psi_\phi$ at each timestep such that Eq.\,\eqref{eq:energies} is satisfied. 
To account for the radiation rapidly dispersing out of previous overdensities, we take this to be spatially homogeneous 
(the same approach was used in N-body simulations of particle dark matter during EMD in Ref.\,\cite{Ganjoo:2023fgg}).  
During EMD, $\left<|\psi_\phi|^2\right> \gg \left<|\psi_a|^2\right>$, and at the end of EMD $\left<|\psi_\phi|^2\right>$ falls towards zero (in practice, we stop evolving the modulus field once it contains a negligible energy density).

We solve Eq.\,\eqref{eq:Schrodinger Poisson} on a lattice of fixed comoving size and lattice spacing with periodic boundary conditions. We use cubic lattices typically containing approximately $1000^3$ points; given the need for multiple simulation runs, this is the largest feasible size with our available computational resources. We use a standard second-order pseudo-spectral algorithm, in which the part of the evolution involving the Laplacian is carried out in momentum space and the potential part in real space \cite{Levkov:2018kau,Schwabe:2020eac}. This approach conserves $\left<|\psi_{a,\phi}|^2\right>$ up to machine precision and does not lead to unphysical growing modes or other instabilities. All resolved modes satisfy $k/R\ll m_a$ throughout, ensuring the validity of the non-relativistic approximation.

The power spectrum of modulus density fluctuations 
$P_{\delta,\phi}$ is defined by
\beq
\left<\delta_{\phi,k} \delta_{\phi,k'}\right> = (2\pi)^3 \delta^{(3)}(k-k')P_{\delta,\phi}(k)~,
\eeq
where $\delta_{\phi,k}$ is the Fourier transform of $\delta_{\phi}(x)=\rho_{\phi}(x)/\bar{\rho}_{\phi}-1$. 
The dimensionless $\phi$ power spectrum $\Delta^2_{\phi}$ is given by
\beq
P_{\delta,\phi} = \frac{2\pi^2}{k^3}\Delta^2_{\phi}(k)~.
\eeq
The axion density power spectra are defined analogously.

We begin simulations during EMD when the temperature of the Universe is $T=100\,\MeV$. At this time, 
the modulus power spectrum is
\beq
\Delta^2_\phi(k)= 0.0023 \left(\frac{k}{R m_a}\right)^4 \anorm{T_{\rm rh}}{10\,\MeV}{8} \anorm{10^{14}\,\GeV}{f_a}{4} ~,
\eeq
corresponding to the growth of sub-horizon modes from an initially scale-invariant spectrum during EMD
\begin{equation}
    \Delta^2_\delta|_{\rm entry}=3.36\cdot 10^{-10}.
\end{equation} 
The initial axion field is taken to be spatially homogeneous. We have checked that, since axion perturbations are driven by the growing $\phi$ perturbations, the difference between starting from a homogeneous axion field and from one with small initial fluctuations is negligible. 
In particular, as shown in Appendix~\ref{app:res}, the axion power spectrum at the end of EMD is insensitive to this choice for $f_a\geq 3\cdot 10^{14}\,\GeV$ with $T_{\rm rh}=20\,\MeV$, as used in our simulations. For smaller $f_a$, more realistic initial conditions may be required for accurate late-time results. 

We make one modification to the physical system: we take the modulus mass (typically tens or hundreds of $\TeV$) to be artificially small while keeping $\Gamma_\phi$ fixed. This ensures that quantum pressure in the modulus field suppresses fluctuations at the lattice scale, 
which would lead to numerical artifacts. Additionally, we give $m_\phi$ an artificial time dependence proportional to $R^{-1/2}$ such that $k_{J,\phi}$ is time-independent during EMD. We also cut off the initial modulus power spectrum at this wavenumber. 
We choose $m_\phi$ such that $k_{J,\phi}$ is at least a factor of about $5$ above the axion mode where there is the most power at the end of EMD, $k_{\rm crit}|_{\rm rh}$. With this condition satisfied, our modification affects only the UV behaviour of the modulus field and does not impact the axion dynamics (specifically, the properties of the axion field at the end of EMD are unaffected). 

Axion self-interactions add a term on the left-hand side of the axion equation of motion in Eq.\,\eqref{eq:Schrodinger Poisson} 
\beq
 - \frac{\lambda}{(2Rm_a)^3}|\psi_a|^2 \psi_a~,
\eeq
with $\lambda\simeq m_a^2/f_a^2$. This is negligible compared to that from the gravitational potential due to the much larger modulus energy density at early times, and it remains unimportant around MRE, consistent with the axion stars lying well below the stability bound.

\subsection{Systematic uncertainties}

There are various sources of systematic errors in our numerical simulations. We have tested these and confirmed that the choices used for our main simulations do not lead to substantial uncertainties.

{\it Finite timestep: }
In simulations, we use the dimensionless time variable $\tilde{t}= m_a \int^t dt'/R(t')^2$, where $R=1$ at $T=100\,\MeV$. Since the timestep $\Delta\tilde{t}$ required for numerical accuracy varies with the gravitational potential, we adapt it throughout simulations as follows. We define
\beq
\Delta \tilde{t}_k = 2\frac{m_a^2}{k_{\rm max}^2}~, \quad 
\Delta \tilde{t}_\Phi =   \frac{2}{R^2 \Phi_{\rm max}} ~,
\eeq
where $k_{\rm max}$ is the highest comoving wavenumber mode allowed by the lattice spacing. 
Schematically, $\Delta \tilde{t}_k$ and $\Delta \tilde{t}_\Phi $ represent the timestep over which the phase
accumulated from the Laplacian and gravitational potential is $\mathcal{O}(1)$  respectively in the Schr\"odinger-Poisson system written in terms of $\tilde{t}$. 
During EMD and radiation domination we use
\beq
\Delta\tilde{t}= \rm{min}\{ \epsilon \sqrt{\Delta \tilde{t}_k \Delta \tilde{t}_\Phi }, 10^3\epsilon \Delta \tilde{t}_k   \}~,
\eeq
where the second possibility ensures the timestep remains finite even during radiation domination when the gravitational potential is negligible. Near MRE, specifically after $\rho_a/\rho_{\rm rad}=10^{-3}$, we use
\beq
\Delta\tilde{t}=\rm{min}\{ \frac{1}{3} \epsilon \sqrt{\Delta \tilde{t}_k \Delta \tilde{t}_\Phi }, 10^2\epsilon \Delta \tilde{t}_k   \}~.
\eeq
Here $\epsilon$ is a numerical parameter. We find that $\epsilon=0.2$ provides sufficient  accuracy. Rerunning simulations with identical initial conditions with 
$\epsilon=0.1$ typically changes the maximum axion energy density at the end of simulations by only a few percent.\footnote{At occasional timesteps the difference is larger, but these errors do not accumulate.} 

{\it Finite box size: }
We take the box size sufficiently large that the smallest wavenumbers $k_{\rm min}$ satisfy $\Delta^2_a(k_{\rm min})\lesssim 0.1$ and $\Delta^2_\phi(k_{\rm min})\lesssim 0.1$ throughout the simulations, so box-scale fluctuations remain perturbative. We have checked that increasing the box size by a factor of two from the values used in our main runs does not alter the form of $\Delta^2_a$ at the end of EMD or after free-streaming. Similarly, provided 
$\Delta^2_a(k_{\rm min})$ remains less than one around MRE, the fraction of axions bound is unaffected by the finite box size. This is consistent with results for vector dark matter produced by inflation \cite{Gorghetto:2022sue} and post-inflationary axions \cite{Gorghetto:2024vnp}, which have similar spectra and dynamics around MRE.

{\it Lattice spacing: }
Increasing the resolution by a factor of two does not affect the axion spectrum at the end of EMD, provided not too much modulus power accumulates at the lattice scale. Around MRE, finite lattice-spacing effects are not significant while the densest solitons, which are the smallest, still contain at least a few lattice points within their cores. We stop simulations when this is no longer satisfied.

\begin{figure}[t!]
    \centering    \includegraphics[width=1.0\columnwidth]{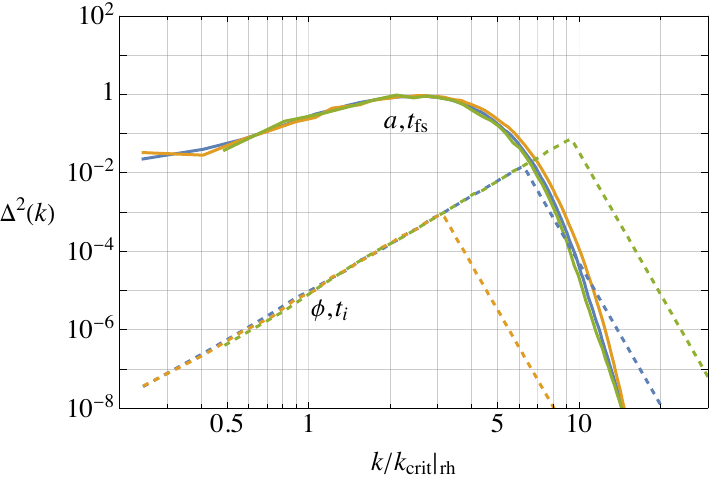}
    \caption{The axion density power spectrum during radiation domination after free-streaming $a,t_{\rm fs}$ (solid lines) for $f_a=10^{15}\,\GeV$ and $T_{\rm rh}=20\,\MeV$, with different modulus Jeans scales and cutoffs. The corresponding initial modulus power spectra are also shown $\phi,t_i$ (dashed).}
    \label{fig:cutoff}
\end{figure}

{\it Modulus mass: }
Since the initial $\Delta^2_\phi\propto k^4$, we cannot place the modulus Jeans cutoff too far above $k_{\rm crit}|_{\rm rh}$ without generating too much UV power, which would lead to numerical artifacts. Typically, we take the cutoff at about $5k_{\rm crit}|_{\rm rh}$. We have checked that the axion power spectrum at the end of EMD is unaffected when this is varied from  $2.5$ to $7.5 k_{\rm crit}|_{\rm rh}$, with the initial modulus power spectrum also cut off at the Jeans scale. 
As a representative example, in Fig.\,\ref{fig:cutoff}, we plot the axion density power spectrum deep into radiation domination, for different modulus Jeans scales spanning this range for $f_a=10^{15}\,\GeV$ and $T_{\rm rh}=20\,\MeV$.\footnote{For the largest cutoff we take a slightly smaller box size to allow a smaller lattice spacing and avoid too much power at the lattice scale.} The late-time axion power spectrum takes approximately the same form in all cases.

\subsection{Further results}

Here we present further results from simulations, for axions with different values of $f_a$.

\begin{figure*}[t]
  \centering
  \begin{minipage}{0.48\textwidth}
    \centering
    \includegraphics[width=\linewidth]{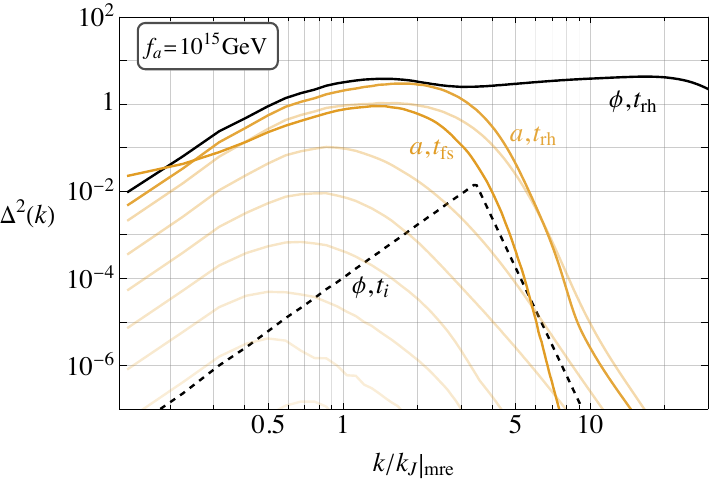}
  \end{minipage}\hfill
  \begin{minipage}{0.48\textwidth}
    \centering
    \includegraphics[width=\linewidth]{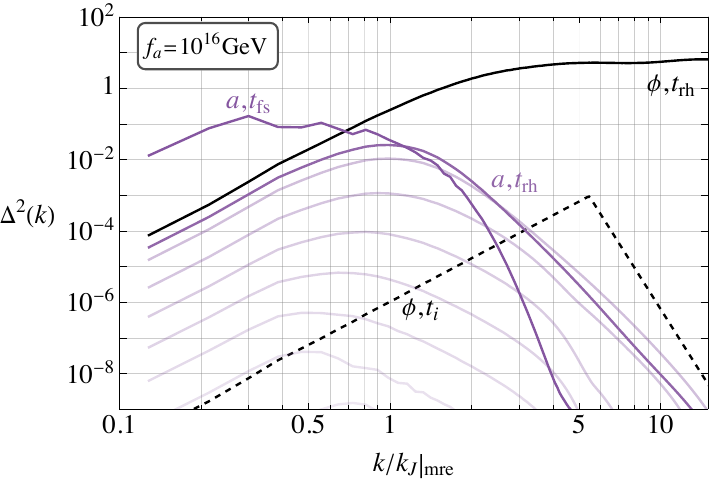}
  \end{minipage}
  \caption{The axion $a$ and modulus $\phi$ density power spectra during EMD, for $f_a=10^{15}\,\GeV$ (left) and $f_a=10^{16}\,\GeV$ (right), both with $T_{\rm rh}=20\,\MeV$. The modulus power spectrum (black) is shown at the initial simulation time $t_i$ and at the end of EMD $t_{\rm rh}$. The axion power spectrum (coloured lines) is shown during EMD (partly transparent lines), at reheating, and after free-streaming $t_{\rm fs}$ during the subsequent radiation domination. The scales are shown in terms of the axion quantum Jeans scale at $\Lambda$CDM matter-radiation equality $k_J|_{\rm mre}$.}
  \label{fig:den_power}
\end{figure*}

In Fig.\,\ref{fig:den_power} we plot the density power spectra of the modulus and axion during EMD, and of the axion after subsequent free-streaming. By free-streaming, we mean the regime in which the gravitational potential is negligible and the axions evolve only under their kinetic term, so they escape the previous gravitational wells.

Results are shown for $f_a=10^{15}\,\GeV$ and $f_a=10^{16}\,\GeV$, both with $T_{\rm rh}=20\,\MeV$ for which $k_{\rm crit}|_{\rm rh}\simeq 0.55\,k_{J}|_{\rm mre}$. For $f_a=10^{16}\,\GeV$, $\Delta^2_a$ remains less than one and increases slightly during free-streaming owing to the peculiar velocities acquired during EMD. The resulting peak lies almost exactly at $k_{\rm crit}|_{\rm rh}$. For $f_a=10^{15}\,\GeV$, $\Delta^2_a$ is mildly nonlinear before the end of EMD, leading to enhanced power at somewhat smaller scales, $k\gtrsim k_{\rm crit}|_{\rm rh}$, at reheating. After free-streaming, the power spectrum relaxes to a peak with $\Delta^2_a\sim 1$ at a scale of about $3\,k_{\rm crit}|_{\rm rh}$. 
The peak is not erased by free-streaming because the axion momentum distribution is peaked around $k_{J}|_{\rm rh}$.

\begin{figure}[t!]
    \centering    \includegraphics[width=1.0\columnwidth]{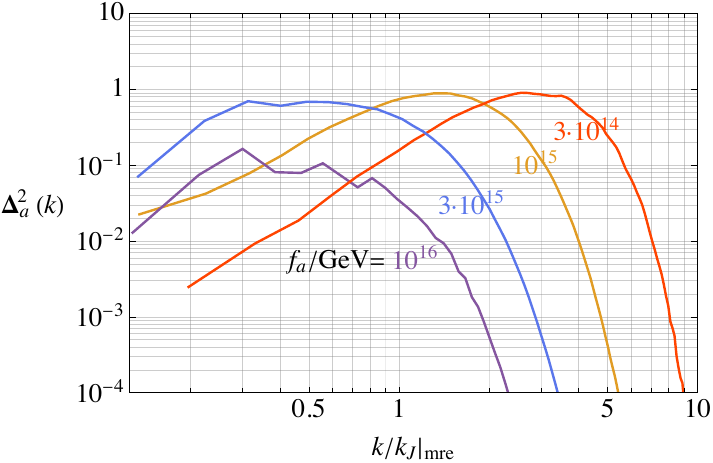}
    \caption{The axion density power spectrum during radiation domination, for different values of $f_a$, with $T_{\rm rh}=20\,\MeV$.   
    }
    \label{fig:denall}
\end{figure}

In Fig.\,\ref{fig:denall} we show the axion density power spectrum after free-streaming, well before MRE, for $f_a=3\cdot 10^{14}\,\GeV$, $10^{15}\,\GeV$, $3\cdot 10^{15}\,\GeV$, and $10^{16}\,\GeV$, all with $T_{\rm rh}=20\,\MeV$. As $f_a$ is decreased from $10^{16}\,\GeV$, the peak approaches the nonlinear regime, as expected from Fig.\,\ref{fig:plot1}. For smaller $f_a$, the modulus power spectrum at $k_{\rm crit}|_{\rm rh}$ becomes increasingly nonlinear. This leads to axion power at progressively smaller scales ($k \gtrsim k_{\rm crit}|_{\rm rh}$), due to axions becoming bound to modulus structures before the modulus decays, shifting the peak of $\Delta^2_a$ after free-streaming to scales a factor of a few above $k_{\rm crit}|_{\rm rh}$.

\begin{figure*}[t]
  \centering
  \begin{minipage}{0.48\textwidth}
    \centering
    \includegraphics[width=\linewidth]{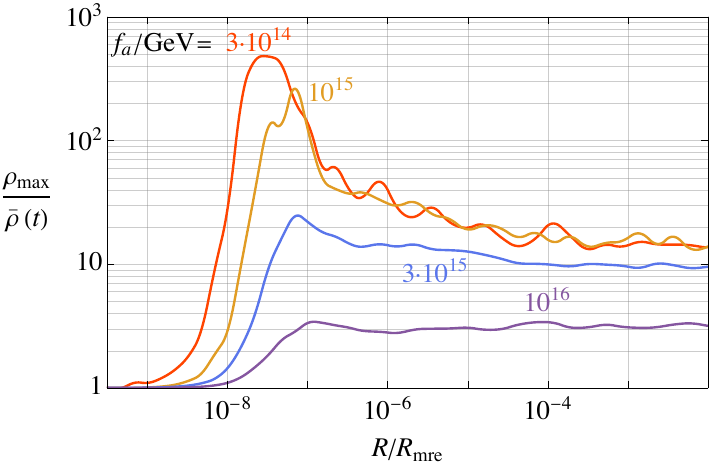}
  \end{minipage}\hfill
  \begin{minipage}{0.48\textwidth}
    \centering
    \includegraphics[width=\linewidth]{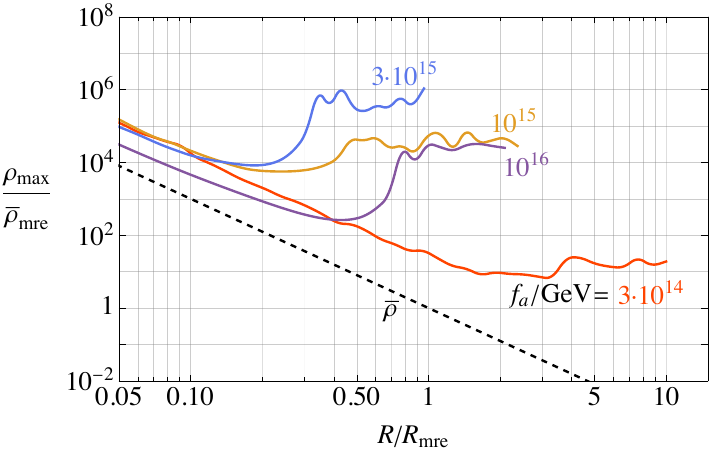}
  \end{minipage}
  \caption{
The maximum axion energy density $\rho_{\rm max}$ in a representative simulation for different $f_a$ with $T_{\rm rh}=20\,\MeV$, as a function of the scale factor normalised to its value at MRE $R_{\rm mre}$. The left panel shows results around EMD, normalised to the evolving mean energy density $\bar{\rho}(t)$. The peak corresponds to axions gravitationally bound to large modulus perturbations. As the modulus decays, the axions free-stream out of the overdensities and relax to a Gaussian random field, corresponding to the plateau in $\rho_{\rm max}/\bar{\rho}(t)$. The right panel shows results at late times, normalised to the mean density at MRE, $\bar{\rho}_{\rm mre}$. The growth and subsequent plateau of $\rho_{\rm max}/\bar{\rho}_{\rm mre}$ corresponds to the formation of axion stars.
  }
  \label{fig:max}
\end{figure*}

In Fig.\,\ref{fig:max}, we plot the maximum axion energy density in a simulation volume for the same parameter points as in Fig.\,\ref{fig:denall}, for representative runs. This is not a strictly physical observable, since it depends on the simulation box size and the particular initial conditions, but it provides a useful indication of the dynamics. Results are shown during EMD, normalised to the time-dependent mean axion density, and around MRE, normalised to the mean axion density at MRE.

During EMD, smaller $f_a$ corresponds to a larger maximum axion energy density, reflecting the larger modulus power spectrum at $k_{\rm crit}|_{\rm rh}$. For $f_a=10^{16}\,\GeV$ and $f_a=3\cdot 10^{15}\,\GeV$, the axion power spectrum remains roughly less than one during EMD. After reheating at $R/R_{\rm mre}\simeq 10^{-7}$, the maximum overdensity remains approximately constant during radiation domination in these cases. Conversely, for $f_a=10^{15}\,\GeV$ and $f_a=3\cdot 10^{14}\,\GeV$, the maximum overdensity decreases after EMD as axions free-stream out of gravitationally collapsed modulus structures.

Shortly before MRE, the largest overdensities in the axion field are about an order of magnitude denser than the background for $f_a=3\cdot 10^{15}\,\GeV$, $10^{15}\,\GeV$, and $3\cdot 10^{14}\,\GeV$. These overdensities collapse earliest for $f_a=3\cdot 10^{15}\,\GeV$, and at successively later times for smaller $f_a$, as indicated by the growth and subsequent saturation of $\rho_{\rm max}/\bar{\rho}_{\rm mre}$. This is because the peak of $\Delta^2_a$ lies at larger $k/k_{J}|_{\rm mre}$ for smaller $f_a$, so quantum pressure delays collapse until later times, when $k_J$ has grown to match the peak scale (since $k_J\propto R^{1/4}$). Meanwhile, for $f_a=10^{16}\,\GeV$, collapse occurs slightly later than for $f_a=3\cdot 10^{15}\,\GeV$, because the initial axion fluctuations are smaller.

\begin{figure*}[t]
  \centering
  \begin{minipage}{0.48\textwidth}
    \centering
    \includegraphics[width=\linewidth]{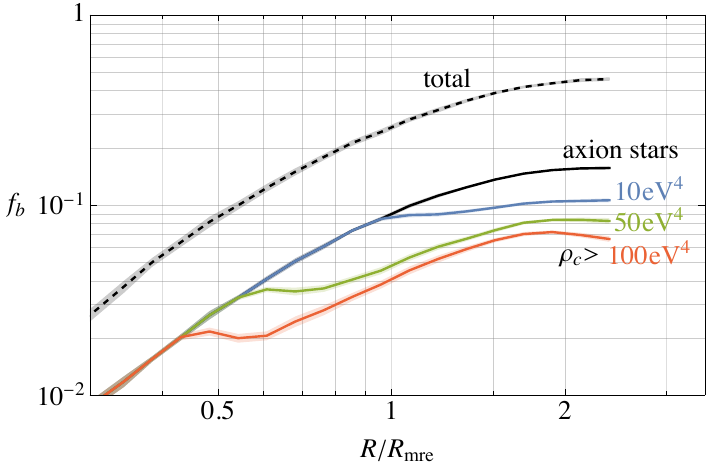}
  \end{minipage}\hfill
  \begin{minipage}{0.48\textwidth}
    \centering
    \includegraphics[width=\linewidth]{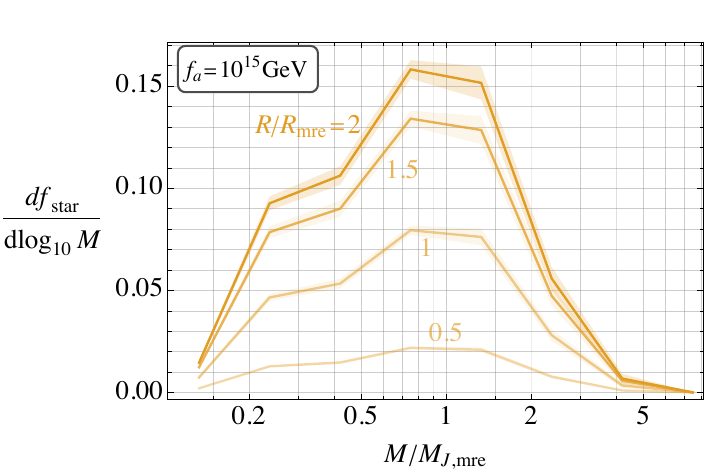}
  \end{minipage}
  \caption{
{\it Left:} 
 The fraction $f_b$ of axions bound in axion stars (black), stars and the surrounding halos (black dashed), and stars with central densities exceeding the given thresholds (coloured) for $f_a=10^{15}\,\GeV$ and $T_{\rm rh}=20\,\MeV$ (with statistical uncertainties). 
 The fraction identified soon after formation is affected by the stars being produced in excited states, but this becomes less significant at late times. 
{\it Right:} The differential fraction of axions bound in stars, $f_{\rm star}$, as a function of axion star mass $M$, evaluated at different times ($R/R_{\rm mre}$), for the same $f_a$ and $T_{\rm rh}$ as the left panel. The masses are normalised to $M_{J,\,\rm mre}$ defined in Eq.\,\eqref{eq:masses2}.
  }
  \label{fig:fall}
\end{figure*}

These results are consistent with Fig.\,\ref{fig:frac} in the main text, which shows that axion stars form later for smaller $f_a$. For $f_a=10^{16}\,\GeV$, the fraction of axions in stars is also slightly smaller (about $5\%$), while the total bound fraction is similar to the other cases. This occurs because the peak of $\Delta^2_a$ lies at slightly smaller $k/k_{J}|_{\rm mre}$ and has a smaller amplitude, so overdensities collapse slightly later, by which time $k_J$ has grown and exceeds the peak scale.

In Fig.\,\ref{fig:fall} (left), we plot the fraction of axions in stars above given central density thresholds for $f_a=10^{15}\,\GeV$ and $T_{\rm rh}$ as a function of scale factor (with stars defined as in the main text). The densest stars, with $\rho_c \gtrsim 10^{2}\eV^4$, form first, followed by progressively lighter, less dense stars at later times 
(corresponding to smaller $\delta_a$ or larger $k_\delta/k_{J}|_{\rm mre}$ in Eq.\,\eqref{eq:masses}).
Even lighter stars are expected to form beyond the range of simulations.

The distribution of axion star masses at different times for $f_a=10^{15}\,\GeV$ and $T_{\rm rh}=20\,\MeV$ is shown in Fig.\,\ref{fig:fall} (right). Specifically, we plot the differential fraction of axions bound in stars of mass $M$. As expected, the majority of bound axions are contained in stars with masses of order $M_{J,\,{\rm mre}}$, as defined in Eq.\,\eqref{eq:masses2}, corresponding to the mass within a Jeans length at MRE. Most of the mass is in stars with central density $\rho_c\sim 10^2\,\bar{\rho}_{\rm mre}$, while the heaviest stars have densities exceeding $10^4\,\bar{\rho}_{\rm mre}$.

For $f_a=3\cdot 10^{14}\,\GeV$, the analogous distribution of axion star masses peaks at slightly smaller values, $M/M_{J,\,{\rm mre}}\simeq 0.5$ (note that the value of $M_{J,\,{\rm mre}}$ depends on $f_a$), while for $f_a=3\cdot 10^{15}\,\GeV$ and $10^{16}\,\GeV$ the peak remains around $M_{J,\,{\rm mre}}$.

We have checked that, when expressed in units of $k_{J}|_{\rm mre}$, the axion power spectrum after free-streaming
for $T_{\rm rh}=10\,\MeV$ and $f_a=1.2\cdot 10^{15}\,\GeV$, $4\cdot 10^{15}\,\GeV$, and $4\cdot 10^{16}\,\GeV$ match those for $T_{\rm rh}=20\,\MeV$ and $f_a=3\cdot 10^{14}\,\GeV$, $10^{15}\,\GeV$, and $10^{16}\,\GeV$ respectively. Specifically, the differences in the peak position and amplitudes are at most a factor of about $2$. 
These values are chosen so that $\Delta^2_a(k_{\rm crit}|_{\rm rh})$, given by Eq.\,\eqref{eq:Dmax}, is matched between pairs of $(f_a, T_{\rm rh})$, indicating that this is the key quantity controlling the overall evolution of the system and the resulting substructure. A detailed exploration of the dynamics across the full range of phenomenologically interesting $(f_a, T_{\rm rh})$ will be carried out in future work.

\nocite{*}

\bibliography{apssamp}

\end{document}